%% file: ms.tex
\documentclass[12pt,preprint]{aastex}

\shorttitle{Stellar Coronal Opacity}

\shortauthors{Matranga et al.}

\begin{document}

\title{Flare X-ray observations of AB Dor: evidence for Stellar Coronal Opacity}

\notetoeditor{the contact email is m.matranga@qub.ac.uk and is the only one which should
appear on the journal version}

\author{M. Matranga\altaffilmark{1},
        M. Mathioudakis\altaffilmark{1},
	H.~R.~M. Kay\altaffilmark{2} and
        F.~P.~Keenan\altaffilmark{1}}

\altaffiltext{1} {Department of Physics and Astronomy, Queen's University Belfast, Belfast, BT7 1NN, Northern Ireland, UK}
\altaffiltext{2} {Mullard Space Science Laboratory, University College London,
	     Holmbury St. Mary, Dorking, Surrey, RH5 6NT, UK}

\begin{abstract}
X-ray spectra of the late-type star AB Dor, obtained with the \emph{XMM-Newton} satellite are analysed. AB Dor was particularly active during the observations. An emission measure reconstruction technique is employed to analyse flare and quiescent spectra, with emphasis on the Fe~XVII~$15-17$~\AA\ wavelength region. The Fe~XVII 16.78~\AA/15.01~\AA\ line ratio increases significantly in the hotter flare plasma. This change in the ratio is opposite to the theoretical predictions and is attributed to the scattering of 15.01~\AA\ line photons from the line-of-sight. The escape probability technique indicates an optical depth of $\approx$ 0.4 for the 15.01~\AA\ line. During the flare, the electron density is $4.4^{+2.7}_{-1.6} \times 10^{10}$ cm$^{-3}$ and the fractional Fe abundance is $0.5 \pm 0.1$ of the solar photospheric value \citep{Grevesse1992}. Using these parameters a path length of $\approx$ 8,000 km is derived. There is no evidence for opacity in the quiescent X-ray spectrum of the star.

\end{abstract}


\section{Introduction}
\label{sect:intro}
Opacity in the solar corona has often been invoked in order to explain the discrepancies in the observed and theoretical spectra of several spectral lines, including O~VIII, Fe~XVII, Ne~IX and Mg~XI \citep{Acton1978,Schmelz1997}. During the process the photon of a resonance line is absorbed by an ion in the ground state, and this excited ion emits a photon at the same wavelength but in a new direction. This process of absorption and re-emission is called resonance scattering. 
The optical depth at line center can be written as \citep{Mitchell1961}: $
\tau_0 \propto \lambda f_{ij} \sqrt{\frac{M}{T}} \frac{n_{\mathrm{i}}}{n_{\mathrm{el}}}
\frac{n_{\mathrm{el}}}{n_{\mathrm{H}}} \frac{n_{\mathrm{H}}}{n_{\mathrm{e}}} n_{\mathrm{e}} l$ where $\lambda$ is the wavelength in \AA, $f_{ij}$ the oscillator strength of the transition, $M$ the mass of the absorbing atom in atomic mass units, $T$ the temperature in K, $l$ the path length in cm and $n_{\mathrm{i}}$, $n_{\mathrm{el}}$, $n_{\mathrm{H}}$, $n_{\mathrm{e}}$ the number densities (in cm$^{-3}$) of ions in the lower level $i$, element, hydrogen and free electrons, respectively.
Some emission lines can have sufficiently large oscillator strengths to lead to high optical depths. Depending on the density and geometry of the emitting region, photons can be scattered out or into the line-of-sight. The method used to study coronal opacity (originally applied to chromospheric line of C~I; \citet{Jordan1967}) relies on pairs of lines from the same element and ionization stage but with different oscillator strengths.
The line with the low oscillator strength, which is used as a reference, is assumed to be optically thin and is often a forbidden transition.
The analysis of solar active regions by \cite{Schmelz1997} revealed statistically significant opacity only in the Fe~XVII line at 15.01~\AA. 

Based on the \emph{EUVE} observations of cool stars, \cite{Schrijver1994} suggested that 
resonance scattering could account for the low line-to-continuum ratio seen in their
coronal spectra. This suggestion was later ruled out by \cite{Schmitt1996} who demonstrated that the count fluctuations in the \emph{EUVE} spectrum suggest that the bulk flux arises from line and therefore the continuum emission had been significantly overestimated (see also \citet{Jordan1996}). Laboratory experiments carried out with the LLNL EBIT-II, have indeed shown a large number of newly identified transitions in the $60-140$ \AA\ wavelength range that add up to a significant flux \citep{Lepson2002}. 

Analysis of \emph{HST} and \emph{FUSE} observations of cool stars have shown significant deviations from the optically thin approximation in the transition region lines of C~III, C~IV, Si~IV and O~VI \citep{Mathioudakis1999,Bloomfield2002}.  
The wealth of high quality X-ray observations from the \emph{Chandra} and \emph{XMM-Newton} missions has allowed the application of line ratio techniques to the coronae of cool stars. 
These techniques were applied to the quiescent coronae of the RS CVn binaries Capella \citep{Phillips2001} and II~Peg \citep{Huenemoerder2001} and revealed no evidence for appreciable opacity. A systematic investigation of opacity effects in 26 late-type stars led to the same conclusion \citep{Ness2003}. More recently \citet{Testa2004} claim the detection of opacity in the O~VIII lines of II~Peg. It therefore seems that, while there is strong evidence for resonance scattering when observing individual solar active regions and flares, these effects are not very well pronounced when observing the emission integrated over the stellar surface.

AB Dor (HD 36705) is a young ($20-30$ Myr), active and rapidly rotating ($v\sin i$ = 90 km s$^{-1}$) K0 dwarf at a distance of 14.9 pc \citep{Unruh1995}.
In the present paper we analyse spectroscopic X-ray observations of AB Dor obtained with \emph{XMM-Newton}. Our particular interest is on the study of resonance scattering in a flare by means of Fe~XVII line ratios.

\section{Observations and Data Reduction}
\label{sect:obs}
\begin{figure}
\begin{center}
\includegraphics[width=17cm]{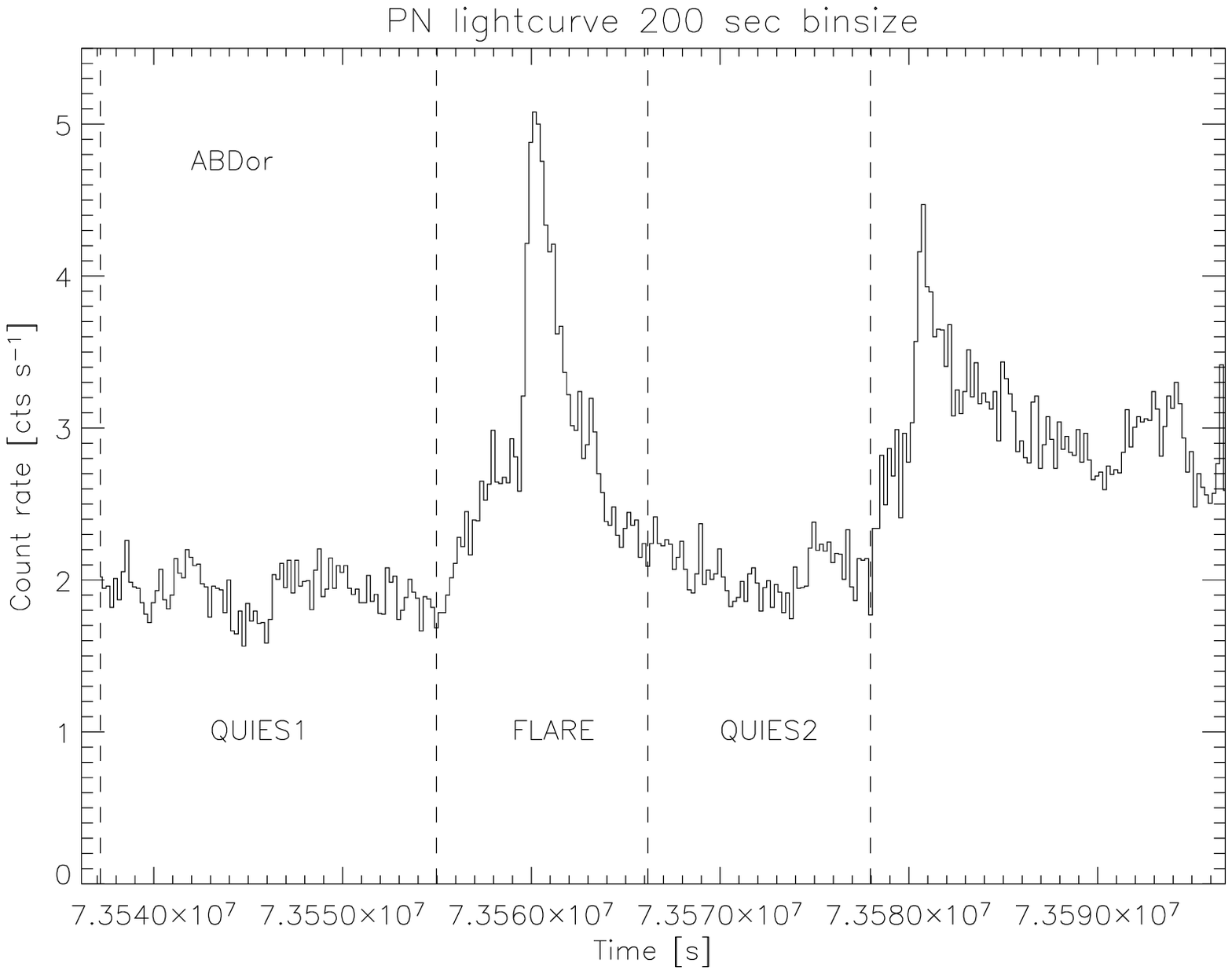}
\caption{The AB Dor light curve derived from EPIC pn using a bin size of 200 sec. During the observation two flares were observed. The flare and quiescent time intervals (FLARE and QUIES1+QUIES2 respectively) used to extract the RGS spectra are shown. Times in seconds since JD 2450814.0 .}
\label{fig:lc_pn}
\end{center}
\end{figure}

The observations presented in this work have been obtained with \emph{XMM-Newton}'s Reflecting Grating Spectrometers RGS1 and RGS2 \citep{denHerder2001}. Data were reduced with the XMM Science Analysis System (SAS) software v5.3.3, updated with the latest calibration files. The light curve derived from the EPIC pn data is shown in Fig. \ref{fig:lc_pn}, where we can see that two flares were observed. 
Using the EPIC pn light curve as a guide, we have extracted RGS spectra for the first flare and quiescent time intervals (FLARE and QUIES1+QUIES2, respectively) as indicated in Fig. \ref{fig:lc_pn}. Based on the analysis of the background light curve extracted from CCD9 of RGS2, time intervals which show high count rates, probably due to proton events, were excluded. The RGS1 and RGS2 spectra are divided into 3400 channels with widths ranging from 0.007 \AA\ to 0.014 \AA\ and cannot be summed directly as their wavelenghts grids do not coincide. Spectra were re-binned into a new grid with a constant width of 0.02 \AA\ before the analysis. Consequently, we have defined a total Line Spread Function (LSF) for the summed spectrum, which is the sum of the single LSFs weighted by the effective area (see \citet{Scelsi2004} for more details).

The analysis performed on the RGS spectra is based on the plasma emission measure distribution (EMD) versus temperature reconstruction technique. We have used the IDL (Interactive Data Language) based software \emph{Package for Interactive Analysis of Line Emission}, PINTofALE v1.5 \citep{Kashyap2000} for this purpose. The analysis can be summarized in the following steps:  

\emph{identification and fit of the lines.} For each spectrum we have identified the
strongest lines using the CHIANTI v4.0 atomic database \citep{Dere1997,Young2003} together with the ionization balance of \citet{Mazzotta1998}; as there is no analytical expression for the total LSF, we have approximated the profile of each line with a Lorentzian having a width fixed at the value which provides the best match with the actual total LSF. \citet{Scelsi2004} have demonstrated that this is a good approximation and that most of the flux  measurements performed in the total spectrum are accurate. It is worth noting that since the wings of the line profile are extended, in the spectrum there is the formation of a ``pseudo-continuum'' which makes measurements of the line flux difficult; multicomponent fitting involving the adjacent transitions was therefore necessary for most of the lines;

\emph{EMD and predicted spectrum.} We have reconstructed the emission measure 
distribution from measured line fluxes using the Markov-Chain Monte Carlo (MCMC) method described by \citet{Kashyap1998}. In order to derive an EMD we have chosen intense lines, i.e. those for which the Lorentzian was a good approximation of the actual LSF and did not fall on the bad pixels of the detector. In the lines selected the Lorentzian profile should reproduce at least 95\% of the actual flux. The chosen lines are reported in Table \ref{tab:lines}. Finally, using the EMD we obtained a predicted spectrum.

\section{Results and Discussion}
\label{sect:results}
\begin{figure}
\begin{center}
\includegraphics[angle=90,width=17cm]{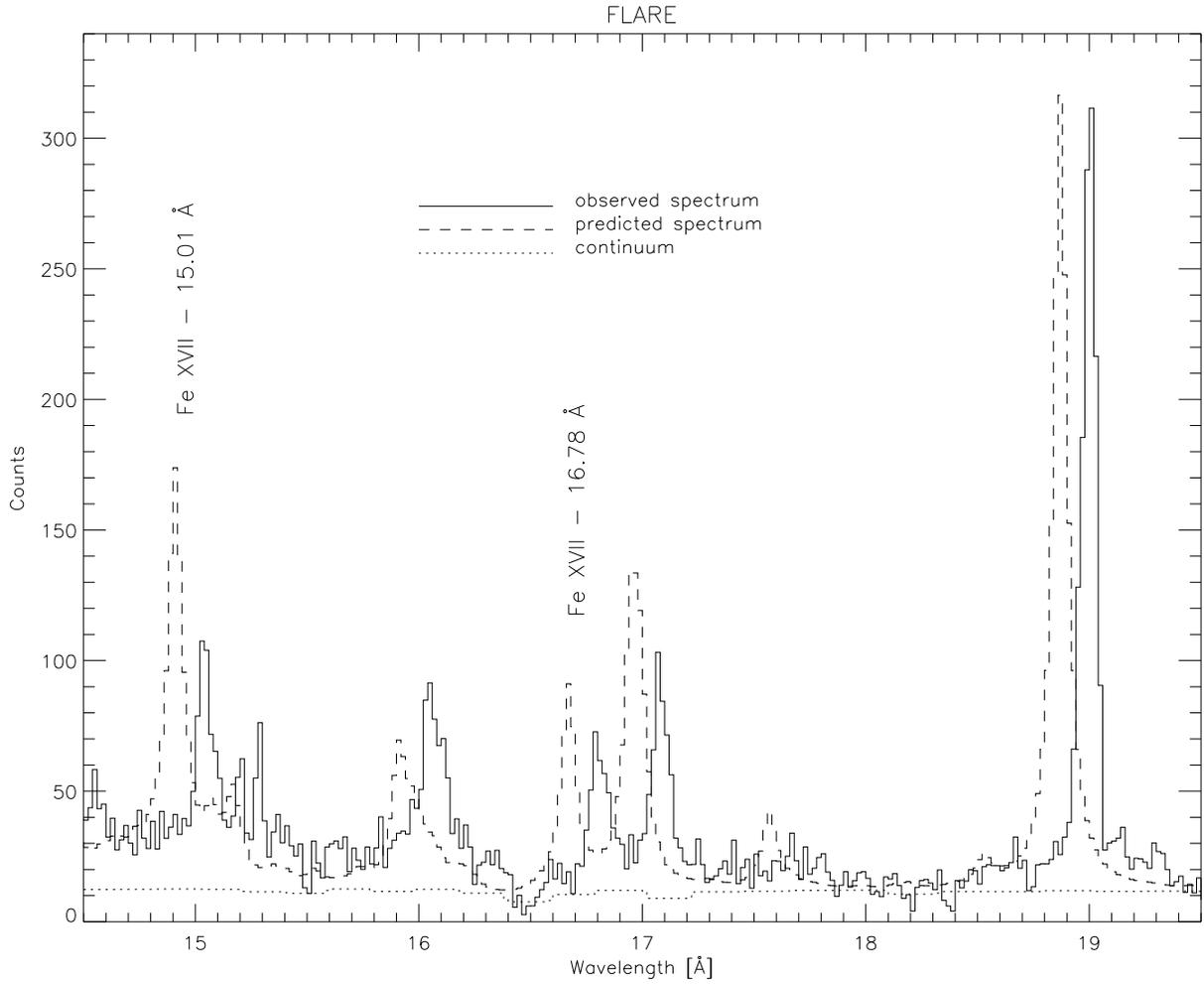}
\caption{Observed spectrum (solid line), predicted spectrum (dashed line) and continuum (dotted line) of the flare. The predicted spectrum has been calculated using the new atomic calculation of \cite{Doron2002}. The wavelengths grid of the predicted spectrum has been shifted by 0.1~\AA\ for clarity. The observed over predicted ratio for the Fe~XVII 15.01~\AA\ line is lower compared to the quiescent state.}
\label{fig:spec_flare}
\end{center}
\end{figure}

\begin{figure}
\begin{center}
\includegraphics[angle=90,width=17cm]{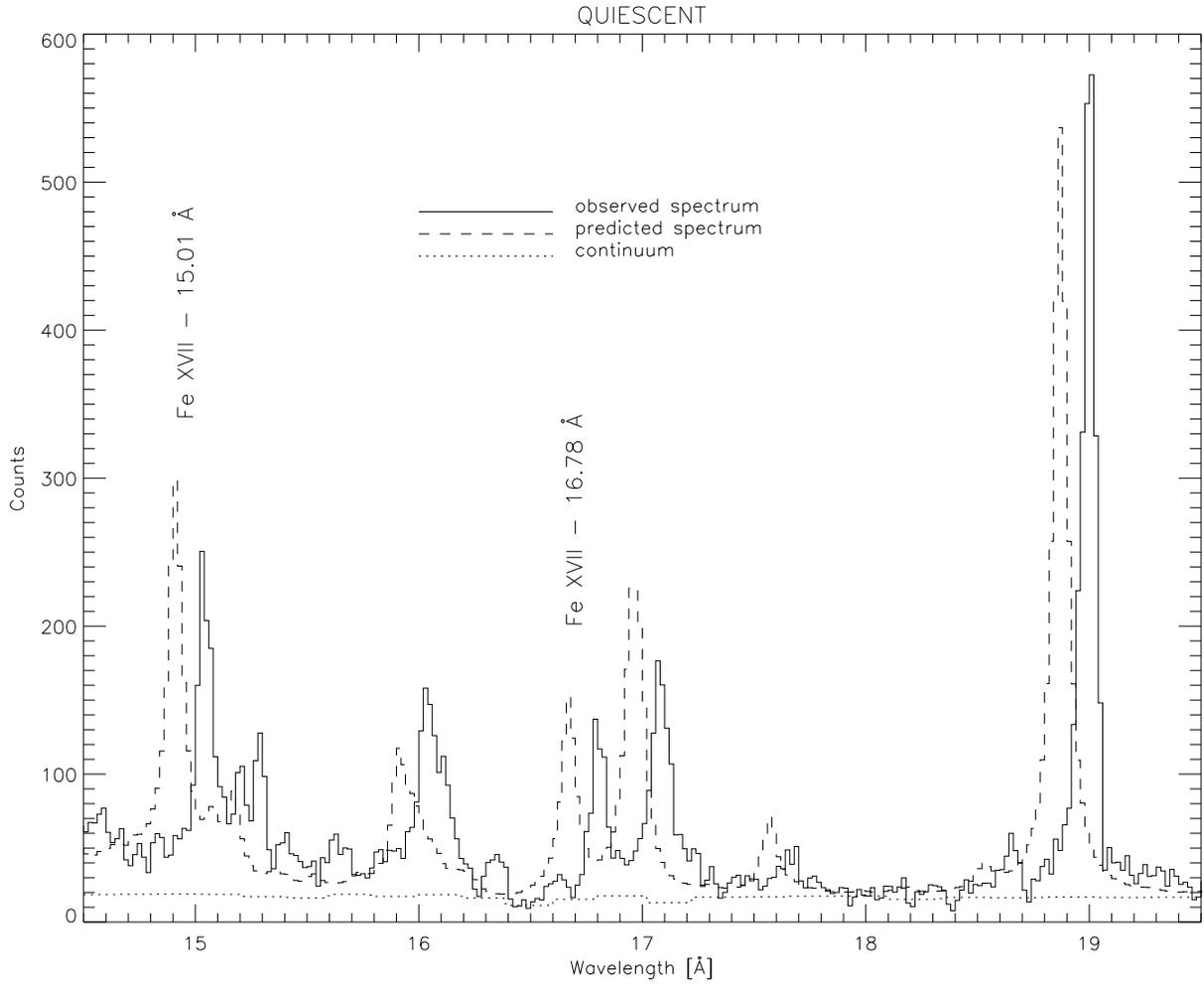}
\caption{The quiescent spectrum of AB Dor. Symbols as in Fig \ref{fig:spec_flare}}
\label{fig:spec_quies}
\end{center}
\end{figure}

In Figures \ref{fig:spec_flare} and \ref{fig:spec_quies} we show the observed and predicted spectra obtained from the analysis of the flare and quiescent data, respectively. While the observed over predicted ratio for the Fe~XVII 16.78~\AA\ line remains the same in the flare and quiescent spectra of AB Dor, the corresponding ratio for Fe~XVII 15.01~\AA\ is reduced significantly in the flare.  

There are some known atomic data problems regarding the Fe~XVII $3d \to 2p$ and $3s \to 2p$ lines located at approximately 15 and 17~\AA . The observed intensity ratios of the $3s \to 2p$ lines (17~\AA) to the $3d \to 2p$ lines (15~\AA) are consistently larger than those calculated by collisional-radiative models. There have been some experimental attempts to explain this discrepancy. \citet{Laming2000} measured the $3s$/$3d$ ratio using the NIST EBIT and found that their data are in agreement with the ratios predicted using a distorted wave collisional model. A similar experiment was carried out by Beiersdorfer et al (2002) at the LLNL EBIT-II: the comparison between their data and theoretical calculations shows that resonant excitation makes a significant contribution to the $3s \to 2p$ lines intensities. This suggests that additional processes have to be taken into account in the interpretation of astrophysical plasmas. \citet{Doron2002} and \citet{Gu2003} have shown that the contribution of radiative recombination (RR), dielectronic recombination (DR), resonant excitation (RE) and inner-shell collisional ionization (CI), in addition to collisional excitation (CE) can rectify some of the disagreements between theory and astrophysical observations. Both studies  show that as the temperature increases the 16.78~\AA/15.01~\AA\ ratio decreases, since the net relative contribution of DR, RR, RE and CI processes for the $3s$ level, compared with collisional excitation, decreases with increasing temperature. These new atomic data of Fe~XVII have been used in our analysis \citep{Doron2002}.

The plasma becomes hotter during the flare. The EMD becomes steeper (EMD$_q\,\propto T^{1.5\pm0.7}$, EMD$_f\,\propto T^{2.7\pm1.0}$ in the temperature region log $T$ = 6.5--7.1) with the minimum shifted to higher temperatures (from log $T_q$ = 6.5 to log $T_f$ = 6.7). We have folded the Fe~XVII 15.01~\AA\ and 16.78~\AA\ line emissivities 
(log $T_{peak}$  = 6.75) with the quiescent and flare EMDs. The contribution function peaks at log $T_{q}$  = 6.8 and log $T_{f}$  = 6.9 for the 
quiescent state and flare respectively. The Fe~XVIII 16.073~\AA\ to Fe~XVII 16.78~\AA\ ratio is often used as a temperature diagnostic, \citep{Phillips1997}, and applied to our observations gives log $T_q$ = 6.6 and log $T_f$ = 6.8. Finally, a fit of the EPIC pn observations of the same flare carried out by \cite{Gudel2001} requires a higher temperature compared to the quiescent state. Although the temperature is therefore higher, the observed 16.78~\AA/15.01~\AA\ ratio increases.

In Table \ref{tab:ratios} we list the observed Fe~XVII 15.01~\AA\ and 16.78~\AA\ line ratios for the quiescent state and flare of AB Dor. The theoretical ratios of \cite{Doron2002} and CHIANTI v4.0 are also listed. While for the quiescent observation the ratio is consistent with the theoretical values, as the temperature increases during the flare, the ratio increases significantly: this observed behavior is opposite to the theoretical one. Using the APED line list \citep{Smith2001} we have folded the emissivities through the emission measure distributions and LSF and found that line blending could account for up to 7\% and 2\% of the 15.01~\AA\ and 16.78~\AA\ line flux respectively. We attribute the increment of the Fe~XVII 16.78~\AA/15.01~\AA\ line ratio that is seen during the flare to the resonance scattering of Fe~XVII 15.01~\AA\ photons out of the line-of-sight.  

Using the line ratio method which has been applied successfully to both stellar and solar spectra \citep{Mathioudakis1999,Bloomfield2002,Schmelz1997}, the reduction in the 15.01~\AA\ line flux allows us to derive the path length of the scattering layer. The photon escape probabilities as a function of optical depth are described in \cite{KastnerKastner1990}. Here we assume an inhomogeneous geometry where the emitters and absorbers are spatially distinct. The ratio between the fluxes of the two emission lines can be written as:

\begin{equation}
\label{eq:ratio}
\frac{F_{16.78}}{F_{15.01}} = \frac{P_{16.78}(\tau_0)}{P_{15.01}(\tau_0)}\, \frac{{{\cal E}}_{16.78}}{{{\cal E}}_{15.01}}
\end{equation}

where $F_{16.78}$ and $F_{15.01}$ are the observed fluxes, $P_{16.78}(\tau_0)$ and $P_{15.01}(\tau_0)$ are the photon escape probabilities and ${{\cal E}}_{16.78}$ and ${{\cal E}}_{15.01}$ are the emissivities for the optically thin (16.78~\AA) and optically thick (15.01~\AA) line respectively. The ratio of the observed fluxes is $\,F_{16.78}/F_{15.01} = 0.71$, while the ratio between the emissivities is $\,{{\cal E}}_{16.78}/{{\cal E}}_{15.01} = 0.54\,$ \citep{Doron2002}. If we assume that $\,P_{16.78}(\tau_0) = 1\,$, from eq.(\ref{eq:ratio}) we have $\,P_{15.01}(\tau_0) = 0.76\,$ which gives an optical depth of $\tau^{15.01}_0 = 0.4\,$  for the Fe~XVII 15.01~\AA\ line. We adopt values of $\tau^{15.01}_0 = 0.4\,$, $f_{ij} = 2.66\,$, $M = 55.8$ amu, $T=10^{6.8}$ K, $n_{\mathrm{i}}/n_{\mathrm{el}} = n_{\mathrm{Fe~XVII}}/n_{\mathrm{Fe}} = 0.63$, $n_{\mathrm{el}}/n_{\mathrm{H}} = n_{\mathrm{Fe}}/n_{\mathrm{H}} = 1.6\times10^{-5}$, $n_{\mathrm{H}}/n_{\mathrm{e}} = 0.8$, $n_{\mathrm{e}} = 4.4\times10^{10}$ cm$^{-3}$. The iron abundance was obtained by matching the continuum of the predicted spectrum to the observed, while the electron density was estimated from the O~VII triplet using the CHIANTI database. Using the above values we derive $l \approx 8\times10^8~cm = 8,000~km $. The uncertainty in the theoretical line ratio is difficult to quantify but should be better understood once the discrepancy between the two laboratory measurements is understood.

Although the Fe~XVII 15.26~\AA/15.01~\AA\ line ratio is a more reliable opacity diagnostic since both lines are populated by collisional excitation from the ground, and the laboratory data for this line ratio are in good agreement, \citep{Laming2000}, the 15.26~\AA\ line is extremely weak and falls on a bad pixel of RGS1. \citet{Testa2004} have found decrements in the O~VIII Ly$\alpha$/Ly$\beta$ ratio in the coronae of very active RS CVn binaries which are interpreted in terms of opacity. We were not able to do a similar measurement here as the Ly$\beta$ line is blended in the RGS spectra. 

\section{Conclusions}
\label{sect:conc}
While it is now accepted that resonance scattering can affect the flux of lines in the solar X-ray spectrum, especially when focusing on small active regions and flares, the results for quiescent stellar coronae remain controversial. Depending on the source geometry, line photons may scatter in or out of the line-of-sight and a net effect is difficult to detect when observing the integrated emission from the stellar surface. Our analysis of the X-ray spectrum of AB Dor has shown a significant increment in the Fe~XVII 16.78~\AA/15.01~\AA\ line ratio during a flare event as compared to the quiescent state. Although the theoretical value of the 16.78~\AA/15.01~\AA\ ratio depends on the adopted atomic model for Fe~XVII, the latest calculations that include both direct and indirect processes predict a decrease in the line ratio with increasing temperature (see \cite{Doron2002} and \cite{Gu2003}). This is opposite to what we observe in the flare. We therefore attribute this change to opacity effects. We demonstrate the potential of coronal opacity as a diagnostic and derive a path length of $\approx$ 8,000 km (0.01 R$_{\star}$) for the coronal scattering layer. This value is two orders of magnitude greater than the path lengths determined in stellar transition regions and falls within the range found for solar coronal loops \citep{Nakariakov2003}.

\acknowledgments
This work is supported by the Particle Physics and Astronomy Research Council (PPARC). M. Matranga would like to thank Queen's University Belfast for a research studentship. F.P.K is grateful to AWE Aldermaston for the award of a William Penney Fellowship. CHIANTI is a collaborative project involving the NRL (USA), RAL (UK), and the Universities of Florence (Italy) and Cambridge (UK). We thank an anonymous referee for his/her comments and suggestions on the paper.

\newpage
\clearpage

\include{tab1}

\newpage
\clearpage

\newpage
\clearpage

\include{tab2}

\newpage
\clearpage

\newpage
\clearpage

\newpage
\clearpage

\end{document}

%% file: tab1.tex
\begin{deluxetable}{l@{~~~}l@{}c@{}l}
\tablewidth{16cm}
\tabletypesize{\scriptsize}
\tablecaption{Lines Used for the Emission Measure Reconstruction\label{tab:lines}}
\tablehead{
\colhead{$\lambda_{{\rm pred}}$\tablenotemark{a}} & 
\colhead{Ion} &
\colhead{$\log T_{{\rm max}}$\tablenotemark{b}} &
\colhead{Blends} 
}

\tablenotetext{a}{~$\lambda_{{\rm pred}}$ is the predicted wavelenght of the transition in \AA.}
\tablenotetext{b}{~$T_{{\rm max}}$ is the maximum temperature of formation of the line in K.}

\startdata
~6.648$^{~f}$ & Si XIII & 7.00 & Si XIII 6.688\\  
~8.425$^{~f}$ & Mg XII  & 7.00 & Mg XII  8.419\\ 
~9.169$^{~f}$ & Mg XI   & 6.80 & \nodata\\  
10.238$^{~q}$& Ne X    & 6.80 & Ne X 10.240\\
11.171$^{~q}$& Fe XXIV & 7.30 & Fe XXIV 11.188\\
11.426$^{~q}$& Fe XXIV & 7.30 & Fe XXII 11.442, Fe XXIII 11.441\\
11.736$^{~f}$& Fe XXIII& 7.20 & Fe XXII 11.767, 11.795, Ni XX 11.832\\
11.767$^{~q}$& Fe XXII & 7.10 & Fe XXII 11.795, Ni XX 11.832\\
12.132$^{~f}$& Ne X    & 6.80 & Fe XXIII 12.175, Fe XXII 12.193, Ne X 12.137, Fe XXI 12.204, Fe XVII 12.123\\
12.285$^{~f}$& Fe XXI  & 7.00 & Fe XVII 12.264\\ 
12.435$^{~fq}$& Ni XIX & 6.90 & Fe XXI 12.398$^{f}$, 12.422$^{fq}$, 12.490$^{q}$, Fe XIX 13.462$^{f}$, 13.464$^{f}$, 13.521$^{f}$, Ne IX 13.447$^{f}$\\
13.447$^{~fq}$& Ne IX  & 6.60 & Fe XIX 13.425$^{q}$, 13.462$^{fq}$, 13.464$^{fq}$, 13.521$^{f}$, Fe XXI 13.507$^{fq}$\\
13.534$^{~f}$ & Fe XIX  & 6.90 & Fe XIX 13.568, Fe XXI 13.574, Fe XXII 13.582, Ne IX 13.553\\
13.699$^{~q}$ & Ne IX   & 6.60 & Fe XIX 13.735, 13.736, 13.795, Fe XX 13.736, Ni XIX 13.779\\	     
15.176$^{~q}$ & O VIII  & 6.50 & Fe XIX 15.198, 15.221, O VIII 15.177\\
16.007$^{~fq}$& O VIII  & 6.50 & Fe XVII 16.006$^{fq}$, Fe XVIII 16.005$^{fq}$, 16.072$^{q}$, Fe XIX 16.027$^{q}$, O VIII 16.006$^{q}$\\
16.165$^{~f}$ & Fe XVIII& 6.90 & Fe XVII 16.240\\
16.778$^{~fq}$& Fe XVII & 6.70 & \nodata\\
17.053$^{~fq}$& Fe XVII & 6.70 & Fe XVII 17.098$^{fq}$\\
18.627$^{~q}$ & O VII   & 6.35 & \nodata\\
18.967$^{~fq}$& O VIII  & 6.50 & O VIII 18.973$^{fq}$\\
20.910$^{~q}$ & N VII   & 6.35 & N VII 20.911\\
21.602$^{~f}$ & O VII   & 6.30 & \nodata\\
21.807$^{~f}$ & O VII   & 6.30 & \nodata\\
22.101$^{~f}$ & O VII   & 6.30 & Ca XVII 22.114\\
24.779$^{~q}$ & N VII   & 6.30 & N VII 24.785\\
28.465$^{~q}$ & C VI    & 6.20 & C VI 28.466\\
28.787$^{~q}$ & N VI    & 6.20 & \nodata\\
30.427$^{~q}$ & S XIV   & 6.50 & S XIV 30.469\\
33.550$^{~q}$ & S XIV   & 6.50 & \nodata\\
33.734$^{~fq}$& C VI    & 6.20 & C VI 33.740$^{fq}$\\
\enddata

\tablecomments{~In the ``$\lambda_{{\rm pred}}$'' column $^{f}$ and $^{q}$ indicate the lines we used for the EM reconstruction for the flare and quiescent observation respectively. In case where the line is selected for both observations, in the ``Blends'' column $^{f}$ and $^{q}$ indicate the blends present with that line for the flare and quiescent spectrum respectively.}

\end{deluxetable}

%% file: tab2.tex
\begin{deluxetable}{cccc|c}
\tablewidth{0cm}
\tabletypesize{\small}
\tablecaption{Observed Line Fluxes and Ratios for the Fe~XVII 15.01~\AA\ and 16.78~\AA\ Lines. Theoretical Ratios are also listed\label{tab:ratios}}
\tablehead{
\colhead{ } &
\colhead{Flux (16.78~\AA)} &
\colhead{Flux (15.01~\AA)} &
\colhead{Obs. Ratio} &
\colhead{Theor. Ratio}\\
\colhead{ } &
\multicolumn{2}{c}{[10$^{-5}$ cnts cm$^{-2}$ s$^{-1}$]} &
\colhead{ } &
\colhead{ }
}

\startdata
quiescent   & $29.7\pm1.5$ & $49.1\pm1.8$ & $0.60\pm0.04$ & $0.57^{{\rm (1)}}$ (600 eV)$\,-\,0.47^{{\rm (2)}}$\\
flare       & $39.1\pm2.6$ & $54.8\pm2.8$ & $0.71\pm0.06$ & $0.54^{{\rm (1)}}$ (800 eV)$\,-\,0.47^{{\rm (2)}}$\\
\enddata
\tablerefs{~$^{{\rm (1)}}$ \cite{Doron2002}; $^{{\rm (2)}}$ theoretical ratios from CHIANTI v4.0}
\end{deluxetable}